**Title: Stability of fault plane solutions for Mw ≥ 4.8 in northern Italy in 2012**

authors: Enrico Brandmayr (1), Fabio Romanelli (1), (2), Giuliano Francesco Panza (1), (2), (3)

(1) Department of Mathematics and Geosciences, University of Trieste, Via Weiss 4, 34127, Trieste, Italy, http://www.dmg.units.it/, enrico.brandmayr@gmail.com, enrico.brandmayr@phd.units.it phone: 040-5582128 fax: 040-5582111

(2) The Abdus Salam International Centre for Theoretical Physics, Strada Costiera 11, 34014 Trieste, Italy

(3) Institute of Geophysics, China Earthquake Administration, Minzudaxuenanlu 5, Haidian District, 100081, Beijing, China

**Abstract**

We propose a critical analysis of the moment tensor solutions of the major seismic events that affected northern Italy in 2012. Inverting full waveforms at regional distance using the non-linear method named INPAR, we investigate period dependent resolution that affects in particular the solutions of shallow events. This is mainly due to the poor resolution of $M_{zx}$ and $M_{zy}$ components of the seismic tensor when inverting signals whose wavelengths significantly exceed the source depth. As a consequence, instability affects both source depth and fault plane solution retrieval, and spurious large Compensated Linear Vector Dipole components arise. The inversion performed at cutoff periods shorter than 20 s reveals in many cases different details of the rupture process, that are not resolved inverting at longer cutoff periods. Thus we conclude that inversion of full waveforms at cutoff period as short as possible should be preferred.

**Introduction**

In the critical analysis of the moment tensor solutions of the recent seismic sequence in Emilia-Romagna, retrieved by the non-linear method named INPAR (Šílený et al., 1992), the reliability of each solution is discussed in terms of accuracy of source time function (STF), percentage of Compensated Linear Vector Dipole (CLVD) component and statistical significance of the retrieved mechanism. The theoretical and practical limits of linear moment tensor retrieval from both body waves (Dufumier, 1996) and surface waves spectra (Dufumier and Cara, 1995) have been widely discussed, in particular for very shallow earthquakes, suggesting to use full waveforms at regional distance for moment tensor inversion. Reduction to zero of the isotropic component does not ensure stable solutions, since it does not affect off-diagonals elements $M_{zx}$ and $M_{zy}$. The poor resolution of these elements, which excite Green functions vanishing at the free surface, causes instability in moment tensor solutions for crustal earthquakes whose source depth is significantly exceeded by the dominant wavelengths of data used in the inversion. As a consequence spurious large CLVD components arise and pairs of solutions with nodal planes showing a 180° rotation around the vertical axis (i.e. $\phi = \phi + 180°$, where $\phi$ is strike angle) are possible, as shown analytically by Henry et al. (2002). The less well determined the $M_{zx}$ and $M_{zy}$ components are, the more likely is the existence of a pair of well-fitting solutions. The conditions for the existence of such a pair are more likely satisfied when dealing with earthquakes close to vertical strike-slip or near-dip-slip mechanism, whose nodal plane dips ~ 45°. Similar conclusions are reached by Bukchin (2006), who proves that the focal mechanism of a seismic source can be uniquely determined from records of surface waves with lengths significantly exceeding the source depth only if the dip angle of one of its nodal planes is sufficiently small.

The real meaning of the retrieved CLVD component is still debated, since it can be even an artifact of the inversion arising from both sparse distribution of stations and inaccurate Earth models (Panza and Saraò, 2000; Henry et al., 2002). The occurrence of sub events with different pure double-couple mechanisms very close in time may lead to the retrieval of a relatively large CLVD component (Guidarelli and Panza, 2007). In this case, different intervals of the source time function can be separately investigated in order to assess how the rupture mechanism changes in time.

**Methodology**

The INPAR non-linear method for the inversion of moment tensor, developed at the Department of Mathematics and Geosciences of the University of Trieste, adopts a point-source approximation. It uses only the dominant part of complete waveforms from events at regional distances (up to 2500 km), thus it maximizes the signal-to-noise ratio. It is particularly suitable in determining shallow event's solutions, due to the possible use of relatively short periods (as short as 10 s). It consists of two steps: in the first, linear, step of the inversion the time functions describing the development in time of the individual components of the moment tensor, namely the moment tensor rate functions (MTRFs), are introduced. Using Einstein summation notation, the $k_{th}$ component of displacement at the surface is the convolution product of the MTRFs and the Green's function spatial derivatives (hereafter Green's functions):

$$u_k(t) = M'_{ij}(t) * G_{ki,j}(t) \qquad (1)$$

The moment rate functions are obtained by deconvolution of the Green's functions from the data. Using the modal-summation technique (Panza, 1985; Florsch et al., 1991; Panza et al., 2000), the synthetic Green's functions are computed at each grid point of a model space defined by a pre-assigned range of possible hypocentral coordinates, and, by interpolation, in the intermediate points, since a bad location of the focus may strongly affect the result of the inversion. The hypocentral location is searched until the difference between synthetic and observed seismograms is minimized. Considering the MTRFs as independent functions in the first step leads to an over-parameterization of the problem which is advantageous for absorbing poor modeling of the structure (Kravanja et al. 1999a). In the non-linear step, the mechanism and the source time function are obtained after factorization of the MTRFs in a time constant moment tensor $m_{ij}$ and a common source time function $f(t)$:

$$\frac{\partial M_{ij}(t)}{\partial t} = m_{ij} \cdot f(t) \qquad (2)$$

routinely assuming the same time dependence for all moment tensor components, i.e. a rupture mechanism constant in time, thus we consider the MTRFs as linearly dependent, taking only their coherent part. The predicted MTRFs are then matched to the observed MTRFs obtained as output of the first step (Panza and Saraò, 2000).

The final solution, obtained by means of a genetic algorithm that allows the estimate of the confidence areas for the different source parameters as well (Šílený, 1998) is consistent with P-wave polarities.

## Results

*Comparison of the solutions.*

The major events ($M_W \geq 4.8$) that affected northern Italy since 25 January 2012, are relocated and their fault plane solution is retrieved, using broadband stations from MEDNET, French, Swiss and Italian Seismic Networks.

The events are chronologically numbered and their source parameters are listed in Table 1. The fault plane solutions retrieved by INPAR, reduced to best double couple, are shown in Figure 1a, compared to TDMT solutions (Fig. 1b, Dreger and Helmberger, 1993), RCMT (Fig. 1c, Pondrelli et al., 2006), both retrieved by the Istituto Nazionale di Geofisica e Vulcanologia (INGV), and CMT-Harvard solutions (Fig. 1d, Dziewonski et al., 1981).

In Figure 1e are shown, for each event, the fault plane solutions and their confidence areas, obtained by INPAR. In order to test the effect of the signals period content on the stability of solutions, the inversion is performed considering cutoff periods varying from 10 to 30 s. In the rightmost column the TDMT solution is shown for comparison. For each event relevant parameters such as source depth, percentage of CLVD and cutoff period are shown.

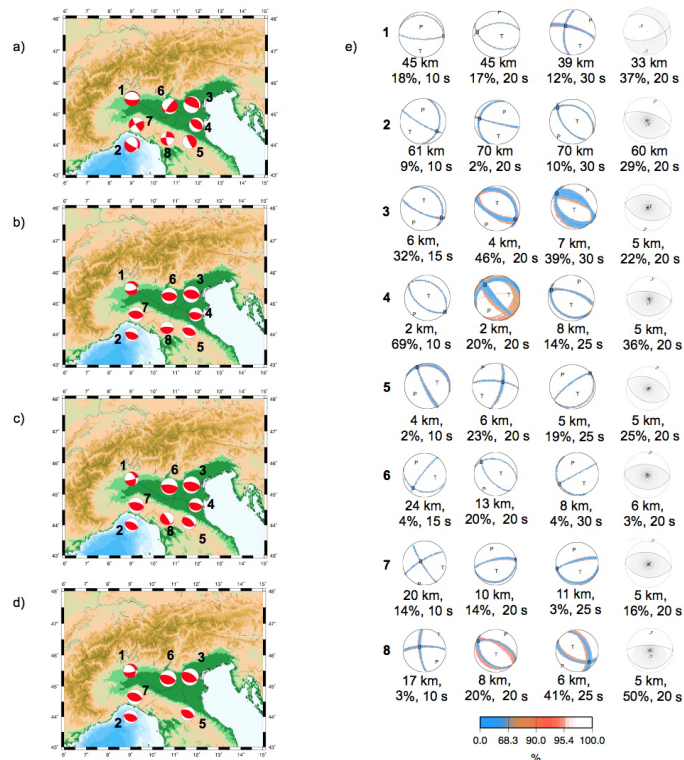

**Figure 1**. Fault plane solutions (best double couple) determined by: a) INPAR, inverting at 10-15 s cutoff period (details in Table 1); b) TDMT-INGV (20 s cutoff period); c) RCMT (30-40 s cutoff

period); d) CMT (40-50 s cutoff period). The beachballs are scaled to magnitude. e) Fault plane solutions determined by INPAR inverting at different cut off periods, with their confidence areas. For each solution, source depth, percentage of CLVD and low pass gaussian filtering period are given. TDMT solutions are reported for comparison in the rightmost column, with the filling of the beachballs denoting the percentage of CLVD component.

The solutions obtained by the inversion at the shortest period (Fig. 1e, first column) are in all cases preferable in terms of confidence areas, percentage of CLVD or both, with the exception of event 1. The most similar solutions between INPAR and results reported by other agencies are found inverting at longer period (Fig. 1e, third column), with the exception of event 1, possibly contaminated by high noise level. Solutions of the events 1, 3 and for 4 are quite in agreement also at shortest period, while event 5 shows a minor strike-slip component not reported by other agencies, that on account of the low percentage of CLVD seems to be reliable.

Event 2, reported as an inverse mechanisms by other agencies, is retrieved by INPAR at the shortest period as a normal mechanism with a minor strike-slip component.

Major differences both in source depth and in the fault plane solutions are found for events 6 to 8 (29 May main shock and aftershocks). INPAR locates these events at a depth ranging from 17 to 24 km and retrieves a significant strike-slip component, while other agencies report shallow and near to 45° dip-slip events, with the exception of event 8 which is reported as a low angle dip-slip both by TDMT and RCMT (Fig. 1b,c). Performing INPAR inversion at longer periods, where the resolution is lost, the strike-slip behavior vanishes and the source depth becomes shallow for all three events.

In the case of the event 3 and 4 a 180° rotation of $\phi$ around the vertical axis is observed in the longer periods INPAR inversion. This can be naturally explained by the poor resolution of $M_{zx}$ and $M_{zy}$ when dealing with shallow events. This poor resolution, as shown in Figure 2, gives rise to pairs of equally probable solutions that satisfy the transformation $\phi \Rightarrow \phi + 180°$ (Henry et al., 2002). The conditions for the existence of such a pair are even more likely satisfied by near-dip-slip earthquake mechanisms with nodal plane dipping ∼ 45°, as in the case of events 3 and 4. The effect of this instability tends to vanish at low (or high) dip angles, which threshold depends on source depth, structure of the medium and spectral content of the inverted signal, but can be generally estimated in 15-20° (Bukchin, 2006).

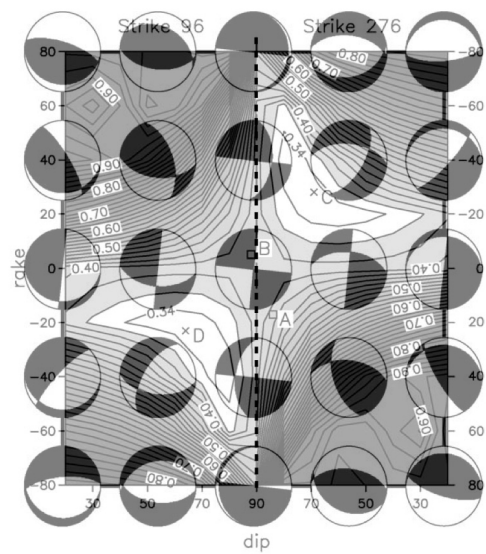

**Figure 2**. Contour plot of the variance ratio for the March 25, 1998 Antarctic earthquake, at strike 96° or 276°, and for a range of dip and rake values, following the Aki and Richards (1980) convention. Points to the left of the vertical dashed line at the center of the figure have strike 96° and rakes corresponding to the left ordinate. Points to the right of the line have strike 276° and rakes corresponding to the right ordinate. The sample moment tensors from the mechanism space, clearly showing the pairs of solutions arising from the two-fold rotational symmetry of the misfit function, are superimposed (modified after Henry et al., 2002).

| N° | Date (yy-m-dd) | UTC Time (hh:mm:ss) | Lon (°) | Lat (°) | Depth (km) | strike (°) | dip (°) | rake (°) | $M_W$ |
|---|---|---|---|---|---|---|---|---|---|
| 1 | 12-1-25 | 08:06:37 | 10.54±.1 | 44.85±.1 | 45±2 | 92 | 74 | -94 | 5.4 |
| 2 | 12-1-27 | 14:53:14 | 10.19±.1 | 44.38±.1 | 61±1 | 120 | 79 | -117 | 5.5 |
| 3 | 12-5-20 | 02:03:53 | 11.23±.1 | 45.00±.1 | 6±1 | 114 | 72 | 78 | 6.0 |
| 4 | 12-5-20 | 03:02:50 | 11.34±.1 | 44.64±.1 | 2±1 | 128 | 54 | 89 | 5.2 |
| 5 | 12-5-20 | 13:18:02 | 11.48±.1 | 44.65±.1 | 4±1 | 153 | 82 | 98 | 5.1 |
| 6 | 12-5-29 | 07:00:03 | 11.21±.1 | 45.00±.3 | 24±14 | 107 | 26 | 158 | 5.7 |
| 7 | 12-5-29 | 10:55:57 | 10.70±.1 | 44.94±.1 | 20±1 | 148 | 81 | 167 | 5.4 |
| 8 | 12-6-03 | 19:20:43 | 11.10±.2 | 44.95±.1 | 17±1 | 269 | 75 | -162 | 4.9 |

**Table 1**. Source parameters of the events retrieved by INPAR inversion at the shortest period (preferred solution).

*Test on STFs and depth constrain.*
To assess the reliability of event's 6 solution, inversion constrained at shallow depths is performed (1-11 km, Fig. 3a). The STF exhibits two peaks and the mechanism solution presents a 43% of CLVD. If the STF is constrained to two sub-intervals, 0-3 s and 3-6 s, the source depth retrieved remains shallow (8 km, quite in agreement with the 6 km retrieved by TDMT) but the still large percentage of CLVD for both solutions obtained considering two separated time intervals for the STFs ("split solutions"), is a strong indicator that the whole solution is unreliable.
Similar results are shown for events 7 and 8 (Figs. 3b and 3c respectively): the double peaked STF is inverted considering two separated time intervals. In both cases the split solutions show percentage of CLVD larger than the whole one, and thus we consider the latter more reliable. Furthermore event 7 shows high instability, with varying source depth and mechanism changing from almost strike-slip to dip-slip between the two time intervals (Fig. 3b), while event 8 shows minor instability (Fig. 3c).

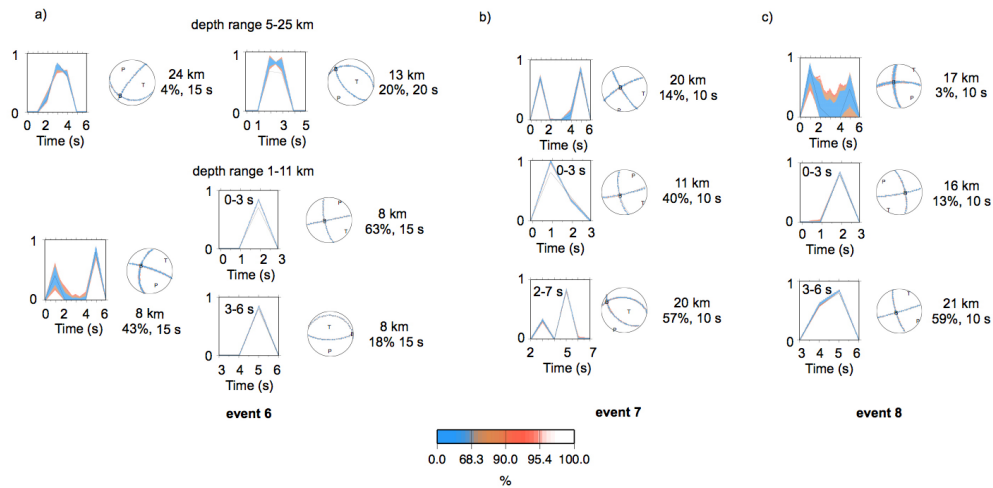

**Figure 3**. a) Top: STF and fault plane solution for event 6 inverted in the depth range 5-25 km. Bottom: results of the inversion of the same event with depth constrained to the range 1-11 km; the double peaked STF is split in two time intervals. The shallow solution seems not reliable due to large percentage of CLVD. b) results of the inversion of event 7 splitting STF in two time intervals. c) results of the inversion of event 8 splitting STF in two time intervals. For each solution source depth, percentage of CLVD and cutoff period are given. The STFs are normalized to unity.

## Conclusions

This paper points out the differences in moment tensor solutions retrieved using signals with different period content. In particular, for shallow crustal earthquakes, instability in the determination of moment tensor solutions arises when signals with a cutoff period longer than 20 s are inverted. Such instability is mainly due to the poor resolution of tensor components $M_{zx}$ and $M_{zy}$, which excite Green functions vanishing at free surface, and it affects both source depth and fault plane solution retrieval. In some cases pairs of equally probable solutions are observed, with a strike angle rotation of 180° around the vertical axis, in particular for near dip-slip events with dip angle ∼ 45°. Thus, in order to assess the reliable fault plane solutions the inversion of full waveforms at periods shorter than 20 s should be preferred.

## Acknowledgments

This research has benefited of the grant "Studio della struttura della crosta e del mantello superiore dell'area mediterranea mediante metodologie sismologiche di inversione non lineare" by Department of Mathematics and Geosciences, University of Trieste. Figures have been plotted using GMT (Generic Mapping Tools; Wessel and Smith, 1995).

## References

Aki, K. and P.G. Richards (1980). Quantitative Seismology, Freeman and Co. New York.
Bukchin B.G. (2006). Specific features of surface radiation by shallow source, Izvestiya. Phys Solid Earth 42(8):712–717.
Dreger, D. S. and D. V. Helmberger (1993). Determination of Source Parameters at Regional Distances with Single Station or Sparse Network Data, Journ. Geophys. Res., 98, 8107-8125.
Dufumier, H. (1996). On the Limits of Linear Moment-tensor Inversion of Teleseismic Body Wave Spectra, Pure and Applied Geophysics, 147, n. 3, 467-482.
Dufumier, H. and M. Cara (1995). On the limits of linear moment tensor inversions of surface wave spectra, Pure and Applied Geophysics 145, 235–257.
Dziewonski, A.M., T.A. Chou and J.H. Woodhouse, (1981). Determination of earthquake source parameters from waveform data for studies of global and regional seismicity. Journal of Geophysical Research, 86, 2825-2852.
Florsch, N., D. Fäh, P. Suhadolc and G.F. Panza (1991). Complete synthetic seismograms for high-frequency multimode SH-waves. Pure and Applied Geophysics, 136, 529-560.
Guidarelli, M. and G.F. Panza (2007). INPAR, CMT and RCMT seismic moment solutions compared for the strongest damaging events (M ≥ 4.8) occurred in the Italian region in the last decade, Proceedings of the National Academy of Sciences called XL, Memoirs of Physical and Natural Sciences, 124°, Vol XXX, t. I, 81-98.

Henry, C., J.H. Woodhouse, and S. Das (2002). Stability of earthquake moment tensor inversions: effect of the double-couple constraint, Tectonophysics, 356, 115-124.

Kravanja, S., G.F. Panza, G.F. and J. Šílený (1999a). Robust retrieval of seismic point source time function. Geophysical Journal International 136, 385-394.

Panza, G.F. (1985). Synthetic Seismograms: the Rayleigh Waves Modal Summation. Journal of Geophysics 58, 125-140.

Panza, G.F., F. Romanelli, and F. Vaccari (2000). Seismic wave propagation in laterally heterogeneous anelastic media: theory and applications to seismic zonation. Advances in Geophysics 43, 1-95.

Panza, G.F. and A. Saraò (2000). Monitoring volcanic and geothermal areas by full seismic moment tensor inversion: are non-double-couple components always artefacts of modelling? Geophysical Journal International, 143, 353-364.

Pondrelli, S., S. Salimbeni, G. Ekström, A. Morelli, P. Gasperini and G. Vannucci (2006). The Italian CMT dataset from 1977 to the present, Phys. Earth Planet. Int., doi:10.1016/j.pepi.2006.07.008, 159/3-4, pp. 286-303.

Šílený, J., G.F. Panza and P. Campus (1992). Waveform inversion for point source moment tensor retrieval with optimization of hypocentral depth and structural model. Geophysical Journal International 109, 259 274.

Šílený, J. (1998). Earthquake source parameters and their confidence regions by a genetic algorithm with a "memory". Geophysical Journal International, 134, 228-242.

Wessel P and W.H.F. Smith (1995). New version of Generic Mapping Tools (GMT) version 3.0 released. Eos Trans AGU 76:329